\documentclass[12pt]{article}
\usepackage{amsmath}
\usepackage{epsfig}
\usepackage{amsfonts}
\usepackage{feynmp}

 \title{Multiboson threshold production by fermion-antifermion pair and hidden symmetry} \author{Joanna Domienik\thanks{supported by the {\L}\'od\'z University grant No.269.},\\ 
 Piotr Kosi\'nski\thanks{supported by KBN grant No. 5 P03B 06021} \\
Department of Theoretical Physics II \\
University of {\L}\'od\'z \\
Pomorska 149/153, 90 - 236 {\L}\'od\'z/Poland.}
 \date{} \begin{document} \maketitle  \begin{abstract} We consider threshold production of arbitrary number of Higgs particles $ \rho$\ by 
fermion-antifermion pair $f \bar{f}$. The nullification of amplitudes for the special case $m_{\rho}=2m_f$\ 
is explained in terms of hidden symmetry of the reduced effective dynamics.
\end{abstract} \newpage
Considerable attention has been paid in the last decade to some aspects of multiparticle 
processes, in particular their threshold behaviour (for a review see \cite{b1}). One of the most 
interesting properties of multiparticle production is the nullification, at the tree level, of
certain threshold amplitudes. There are, basically, two kinds of such processes. First, one can
consider the "hard-soft" scattering, where two initial particles of opposite momenta scatter to 
produce an arbitrary number of final bosons, all at rest. It has been shown that, for a quite 
wide variety of processes, the relevant threshold amplitudes vanish provided the parameters of the 
lagrangian are related in a certain way \cite{b2} $\div$\ \cite{b9}. Typically, these relations take 
the form of "quantization conditions" and the integer  number entering there gives an upper
bound for the number of final bosons which can be produced. For example, consider the production of
$n \; \Phi_2$- particles at rest by two $\Phi_1$- particles of opposite momenta through the
interaction described by 
\begin{eqnarray}
V(\Phi_1, \; \Phi_2)=\frac{g}{2}\Phi_1^2\Phi_2^2+\frac{\lambda}{4!}\Phi_2^4\; ; \label{w1}
\end{eqnarray}
if $g=\frac{N(N-1)}{6}\lambda$, all amplitudes $2\rightarrow n$\ vanish for $n \geq  2N$. In particular,
for $N=2$\ the only nonvanishing amplitude is $2\rightarrow 2$; it is given by the elementary vertex 
$\Phi_1^2\Phi_2^2$. Other examples cover the multiparticle production by vector bosons and 
fermions (see below).

The second class of processes exhibiting nullification phenomena consist of the "soft-soft"
ones where all particles, both initial and final, are at rest. In two nice papers \cite{b10} 
\cite{b11} Libanov et al. considered specific model exhibiting nullification of "soft-soft" amplitudes. 
They have shown that this 
property is related to the integrability of the reduced classical dynamical system obtained by 
neglecting space dependence in the original theory. Their argument was based on the relation 
between the appearance of resonances in the course of perturbative solution of dynamical 
equations for reduced model and nontriviality of the corresponding tree amplitudes. It has been proven in \cite{b11} 
that such resonances are exluded provided the
reduced theory has additional symmetry. The arguments presented in Ref. \cite{b11} were subsequently supported by diagrammatic \cite{b12} and functional
\cite{b13} proofs of the relevant Ward identities expressing the underlying symmetry at the tree level. Moreover, they allowed to construct other models with
vanishing soft-soft amplitudes \cite{b14}.\\
One can pose the question whether the nullification of hard-soft amplitudes can be also explained as a consequence of some 
symmetry. Recently, the present authors proposed a partial solution to this problem \cite{b13} (see also \cite{b15}). 
Consider the $2\rightarrow n$\ hard-soft process; we assume that both inital and final particles are scalars and , 
furthermore, they are of different kind. Moreover, assume that the interaction conserves the lines of 
initial particles (i.e. the interaction   is  even   function of the corresponding field ). 
Then the nonvanishing threemomentum $\vec{p} $\
flows only through the continuous  line of propagators connecting both initial lines. One can immediately conclude that our 
amplitude coincides, up to irrelevant factors, with the one for the process $2\rightarrow n$\ with all, both initial and final, 
particles at threshold and the mass $m$\ of the initial particles replaced  by $M=\sqrt{m^2+\vec{p}^2}$. The vanishing
of the resulting soft-soft amplitude is then the result of integrability of reduced system (Garnier system for the example
considered in Ref. \cite{b13}). In particular, the $N=1$\ case of the model (\ref{w1}) can be treated in this way. 

In this
letter we consider another, more interesting example of the symmetry underlying the nullification phenomenon for the production
of scalar bosons by fermion-antifermion pair. Due to the spin structure of fermion propagator the reduced model is more 
complicated: it is not sufficient to modify the mass only. Still, one can show that the reduced dynamics exhibits nontrivial
symmetry.

Our starting point is the lagrangian
\begin{eqnarray}
L=i\bar{\psi}\gamma^{\mu}\partial_{\mu}\psi+\frac{1}{2}(\partial_{\mu}\Phi \partial^{\mu}\Phi +m^2\Phi^2)-\frac{\lambda}{4!}
\Phi^4-g\Phi \bar{\psi}\psi, \label{w2}
\end{eqnarray}
which is a toy-model describing the fermion mass generation via Yukawa coupling with Higgs field. Define the physical field
$\rho$\ by
\begin{eqnarray}
\Phi \equiv \rho +v\equiv \rho +\sqrt{\frac{6m^2}{\lambda}}\;;\label{w3}
\end{eqnarray}
then
\begin{eqnarray}
L=\bar{\psi}(i\gamma^{\mu}\partial_{\mu}-M)\psi+\frac{1}{2}(\partial_{\mu}\rho \partial^{\mu}\rho -m^2_{\rho}\rho^2)-
\frac{\lambda v}{3!}\rho^3-\frac{\lambda}{4!}\rho^4-g\rho\bar{\psi}\psi \;;\label{w4}
\end{eqnarray}
here $M=gv,\;m_{\rho}=\sqrt{2}m$\ are fermionic and bosonic masses, respectively.

We want to calculate the amplitude for $f\bar{f}\rightarrow n\rho$\ process with all final $\rho$-particles at rest.
This can be done diagrammatically \cite{bA} or by calculating the relevant fermionic propagator in external scalar field 
\cite{b6} or, else, by Feynman wave function method \cite{b2}, \cite{b16}, \cite{b15}. We adopt here the last approach;
accordingly, we first solve the field equation for $\rho$\ (assuming $\psi =0=\bar{\psi}$) under the condition
\begin{eqnarray}
\rho \mid_{\lambda =0}=\frac{\beta e^{im_{\rho}t}}{\sqrt{(2\Pi )^32m_{\rho}}}\equiv z_0e^{im_{\rho}t}\equiv z(t)\label{w5}
\end{eqnarray}
The solution reads \cite{b17}
\begin{eqnarray}
\rho (t)=\frac{z(t)}{1-\frac{z(t)}{2v}}\label{w6}
\end{eqnarray}
Then we look for the Feynman wave function obeying
\begin{eqnarray}
&&(i\gamma^{\mu}\partial_{\mu}-g(v+\rho (t)))\psi_{ps}(x)=0 \label{w7} \\
&&\psi_{ps}(x)\mid_{g=0\atop gv=M}=\sqrt{\frac{M}{(2\pi)^3Ep}}u(p,s)e^{-ipx}\label{w8}
\end{eqnarray}
Having found $\psi_{ps}(x)$\ one can compute the generating functional for tree-level $f\bar{f}\rightarrow n\rho$\
amplitudes by reducing the initial antifermion through LSZ formula. Puting
\begin{eqnarray}
\bar{v}_{p's'}(x)=\sqrt{\frac{M}{(2\pi )^3E_{p'}}}\bar{v}(p',s')e^{-ip'x}\label{w9}
\end{eqnarray}
we arrive at the following form of generating functional \cite{b15}, \cite{b16}
\begin{eqnarray}
-i\int d^4x\bar{v}_{p's'}(x)(i\gamma^{\mu}\partial_{\mu}-M)\psi_{ps}(x) \;;\label{w10}
\end{eqnarray}
succesive derivatives with respect to $\beta$\ at $\beta =0$\ produce the relevant amplitudes. Carefully isolating pole
contribution one finds \cite{b15}
\begin{eqnarray}
&&A(f\bar{f}\rightarrow n\rho )=\label{w11} \\
&&\frac{-4\pi iM\delta(2E_p-nm_{\rho})\delta^{(3)}(\vec{p}+\vec{p}')}{(2\pi)^{3n/2}
(2m_{\rho})^{n/2}(2v)^n(n-1)!} \frac{\Gamma(n+\frac{2M}{m_{\rho}})\Gamma(n-\frac{2M}{m_{\rho}})}{\Gamma(
\frac{2M}{m_{\rho}})\Gamma (1-\frac{2M}{m_{\rho}})}
\vec{v}(p',s')u(p,s)\nonumber
\end{eqnarray}
It follows from the above formula that the amplitude vanishes for all $n>N$\ provided $\frac{2M}{m_{\rho}}=N,\;N$\ 
being an integer. In particular, for $N=1$, i.e. $m_{\rho}=2M$, all amplitudes vanish except $A(f\bar{f}\rightarrow \rho )$;
note that in the latter case all particles are at rest and the amplitude is given by direct Yukawa coupling. Actually, for
any integer $N$\ the amplitudes vanish for all $n$; however, for $n<N$\ this due to the energy conservation, while for 
$n=N$-due to the spinor structure of (\ref{w11}). So, finally, for $n\leq N$\ the vanishing is implied by standard symmetries
while for $n>N$\ the more sophisticated arguments should enter the game. We want to offer the explanation, based on symmetry,
for nullification at $N=1$. As in the purely bosonic case we compare our amplitude with the soft-soft one for suitably 
modified theory. To this end note first that $\psi_{ps}(x)$\ generates matrix elements of the fermionic field between
initial state containing one particle carrying the momentum $\vec{p}$\ and the final one containing an arbitrary number of
particles at rest. Therefore,
\begin{eqnarray}
\psi_{ps}(x)=\psi_{ps}(t)e^{-ip_xx^k}\label{w12}
\end{eqnarray}
Taking into account this relation and carefully inspecting the algorithm leading to the formula (\ref{w11}) we easily
conclude that the relevant amplitudes are equal, up to irrelevant factor, to the soft-soft ones for the theory described by the
lagrangian
\begin{eqnarray}
&&L_{eff}=\bar{\psi}(i\gamma^{\mu}\partial_{\mu}+(\gamma^kp_k-M))\psi+\frac{1}{2}
(\partial_{\mu}\rho \partial^{\mu}\rho -m^2_{\rho}\rho^2)+\nonumber \\
&&-\frac{\lambda v}{3!}\rho^3-\frac{\lambda}{4!}\rho^4-g\rho \bar{\psi}\psi \;;\label{w13}
\end{eqnarray}
Again, as in the scalar case, this conclusion is justified by the fact that the external momentum $\vec{p}$\ flows through
the continuous line of fermionic propagators connecting both initial lines. In order to find $2\rightarrow n$\ soft-soft
amplitude for the theory described by the lagrangian (\ref{w13}) we proceed in a standard way \cite{b2}, \cite{b16}, \cite{b15}.
First, neglecting the space-dependence in $L_{eff}$\ we arrive at the reduced fermionic-bosonic system of finite degrees of
freedom given by
\begin{eqnarray}
&&L_{\kappa d}=\bar{\psi}(i\gamma^{0}\partial_{t}+(\gamma^kp_k-M))\psi+\frac{1}{2}
(\dot{\rho}^2-4M^2\rho^2)+\nonumber \\
&&-\frac{\lambda v}{3!}\rho^3-\frac{\lambda}{4!}\rho^4-g\rho \bar{\psi}\psi \;,\label{w14}
\end{eqnarray}
where the relation $m_{\rho}=2M$\ has been already taken into account. To find the relevant 
purely soft amplitudes we solve the equation of motion following from the lagrangian (\ref{w14}) together with the boundary
condition (\ref{w5}) and
\begin{eqnarray}
&&\psi (t)\mid_{g=0}=\zeta \sqrt{\frac{M}{(2\pi)^3E_p}}u(p,s)e^{-iE_pt}\label{w15}\\
&&\bar{\psi} (t)\mid_{g=0}=\bar{\zeta} \sqrt{\frac{M}{(2\pi)^3E_p}}\bar{v}(-p,s')e^{-iE_pt}\label{w16}
\end{eqnarray}
where $\zeta,\;\bar{\zeta}$\ are anticommuting parameters. The solution is unique provided we take into account 
additional conditions following from the tree-graph interpretation of the solution \cite{b15}. Now,
\begin{eqnarray}
\psi (t)=\psi (t\mid \zeta,\bar{\zeta},\beta )\label{w17}
\end{eqnarray}
generates matrix elements of fermionic field between the initial states of fermion at rest and final ones describing bosons
at rest (actually, due to the fermion number conservation and Pauli exclusion principle $ \psi (t)$\ depends only (linearly) 
on $ \zeta $\ ). 
Applying LSZ we get the generating function for matrix elements for purely soft processes $ f\bar f\rightarrow n\rho $. 
In analogy with scalar
case we expect (\ref{w14}) to exhibit some hidden symmetry. Indeed, it is straightforward although tedious to check that it
posses the integral of  motion of the form
\begin{eqnarray}
&&F=\frac{\vec{p}^2}{M^2}(\dot{\rho}^2+M^2\rho^2(2+\frac{g\rho}{M})^2)-\frac{ig\dot{\rho}}{M^2}\bar{\psi}\gamma^0\gamma^k
p_k\psi+\nonumber \\
&&-\frac{g\rho}{M}(2+\frac{g\rho}{M})\bar{\psi}\gamma^kp_k\psi-\frac{\lambda}{48M^2}(\bar{\psi}\psi )^2\label{w18}
\end{eqnarray}
$F$\ generates the following symmetry transformations
\begin{eqnarray}
&&\rho' =\rho +\varepsilon \left(\frac{2\vec{p}^2}{M^2}\dot{\rho}-\frac{ig}{M^2}\bar{\psi}\gamma^0\gamma^kp_k\psi \right) \label{w19}\\
&&\psi'=\psi + \nonumber\\
&&+\varepsilon \left( -\frac{g\dot{\rho}}{M^2}\gamma^kp_k\psi +\frac{ig\rho}{M}(2+\frac{g\rho}{M})
\gamma ^0\gamma^kp_k\psi+\frac{i\lambda}{24M^2}(\bar{\psi}\psi )\gamma ^0\psi \right) \label{w20}\\
&&\bar{\psi}'=\bar{\psi} +\nonumber\\
&&+\varepsilon \left(-\frac{g\dot{\rho}}{M^2}\bar{\psi}\gamma^kp_k-
\frac{ig\rho}{M}(2+\frac{g\rho}{M})
\bar{\psi}\gamma^kp_k\gamma ^0-\frac{i\lambda}{24M^2}(\bar{\psi}\psi )\bar \psi \gamma ^0 \right ) \label{w21}
\end{eqnarray}
$\varepsilon$\ being an infinitesimal commuting parameter. The symmetry transformations (\ref{w19}) $\div$\ (\ref{w21}) have 
desirable property \cite{b11}, \cite{b13}: the coupling-independent part is linear in dynamical variables. Therefore, one 
can repeat the arguments implying the amplitude nullification. In particular, taking into account that (\ref{w19}) $\div$\ 
(\ref{w21}) induce the transformations of $\zeta , \;\bar{\zeta}$\ and $\beta$\ of the form
\begin{eqnarray}
&&\beta '=\beta (1+4i\frac{\varepsilon \vec{p}^2}{M})\nonumber \\
&& \zeta '=\zeta \nonumber \\
&&\bar{\zeta}'=\bar{\zeta} \nonumber
\end{eqnarray}
we obtain, following the arguments presented in Ref. \cite{b13}, the following Ward identity 
\begin{eqnarray}
\frac{n \vec{p}^2}{M}A(f\bar{f}\rightarrow n \rho)=0\label{w22}
\end{eqnarray}
The nontrivial amplitude is possible only for $\vec{p}=0$; then, by energy conservation, $n=1$\ and the amplitude is simply 
$-ig$\ (but also vanishes when multiplied by fermionic wave functions).

As it has been already noticed in \cite{b13}, there is an analogy with the Coleman-Mandula theorem \cite{b18}. The conserved
charge contains "space-time indices" (the time derivative here) which implies additional constraints on amplitudes (in our
case force them to vanish). In the limit $\vec{p}=0$\ the symmetry becomes purely internal (basically-fermion number 
conservation) and put much weaker restrictions on the form of amplitudes. 

It would be nice to find similar explanation
for $N>1$. Then the amplitudes vanish either due to the properties of hipergeometric function (for $n>N$) or by energy
conservation (for $n<N$); for $n=N$\ one deals with purely soft amplitude vanishing, as mentioned above, due to the properties
of fermionic wave functions. Therefore, the hidden symmetry we are looking for should imply Ward identities responsible for
nullification of amplitudes for $n>N$. This seems to be more complicated than in the case considered here.

\end{document}